\documentclass[twocolumn,showpacs,amsmath,amssymb,floatfix,pre]{revtex4}
\usepackage{graphicx}
\usepackage{epstopdf} 

\begin{document}
\draft

\title{The Glass Transition Temperature of Water: A Simulation Study}

\author{Nicolas Giovambattista$^1$, C. Austen Angell$^2$,
 Francesco Sciortino$^3$ and H. Eugene Stanley$^1$}

\address{$^1$Center for Polymer Studies and Department of Physics,
  Boston University, Boston, Massachusetts 02215.}
\address{$^2$Dept. of Chemistry and Biochemistry \\
  Arizona State University, Tempe, AZ 85287 USA}
\address{$^3$Dipartimento di Fisica, Istituto Nazionale per la Fisica della
Materia,\\
and I.N.F.M. Center for Statistical Mechanics and Complexity,\\
Universit\`{a} di Roma La Sapienza, Piazzale A. Moro 2, I-00185 Roma, ITALY}

\date{Dec. 24th, 2003}


\begin{abstract}
We report a computer simulation study of the glass transition for water.
To mimic the difference between standard and hyperquenched glass, we
generate glassy configurations with different cooling rates 
and calculate the $T$ dependence of the
specific heat on heating. The absence of crystallization phenomena
allows us, for properly annealed samples, to detect in the specific heat 
the simultaneous presence of a weak pre-peak (``shadow transition''), and an
 intense glass transition peak at higher temperature.
 We discuss the implications for
the currently debated value of the glass transition temperature of water.
We also compare our simulation results with the 
Tool-Narayanaswamy-Moynihan phenomenological model.
\end{abstract}

\maketitle



\bigskip\bigskip

\noindent

Much recent research has focused on the properties of glassy
water, the most common form of water
 in the universe, which can exist in
more than one distinct amorphous form\cite{debene,universe,angellRev}.
The conversion between different glass
structures, the different routes producing glass structures, and the relation
between the liquid and the glass phases are under active debate. 

A particularly relevant aspect of this debate concerns the
identification of the glass transition temperature $T_g$ at ambient
pressure and the magnitude of the associated jump of the specific heat,
an issue which has relevance also for determining the fragility of
water.  Extrapolation of $T_g$ in binary aqueous solutions, in the limit
of vanishing solute concentration, provides the estimate 
$T_g \approx 136$~K \cite{binary}.
  Early differential scanning calorimetry (DSC)
studies report conflicting results.  Some experiments detect the glass
transition \cite{did} but others do not \cite{didnot}. An
exothermal peak in the specific heat of properly-annealed hyperquenched
water supports the estimate $T_g \approx 136$~K\cite{Tg136}, with a specific
heat jump of $1.6-1.9$~J/mol/K.  This $T_g$
value\cite{hallbrucker1,itoAngell} has been recently debated
\cite{Tg165,Johari2,angell}.
  It has been suggested\cite{angell} that the small peak
measured in Ref.~\cite{Tg136} is a pre-peak typical of annealed
hyperquenched samples preceding the true glass transition
located at $T_g \approx 165$~K.  Assigning $T_g \approx 165$~K would
explain many of the puzzles related to the glass transition in water
\cite{itoAngell,Tg165,angell}.  Unfortunately, the $T_g \approx 165$~K
proposal can not be experimentally tested due to the homogeneous
nucleation of the crystal phase at $T_\times \approx 150$~K.

Here we report a numerical study of the temperature dependence of the
specific heat across the glass-to-liquid transition for the extended
simple point charge (SPC/E) model for water.  We analyze the effects
both of the cooling rate and of annealing (``aging'') before heating the
glass, since both effects are important for determining $T_g$
\cite{johariWaterAB,tucker}, and both effects have been studied
extensively in many
materials\cite{moynihan,andMore}.
 Numerical studies are particularly suited since crystallization does
 not take place on the time scale probed in simulations.  With an
 appropriate choice of the heating and cooling rates to mimic the
 experimental conditions, we show that both the shadow and the glass
 transition peaks can be resolved in the same heating scan.  Finally, we
 compare the simulation results with the Tool-Narayanaswamy-Moynihan (TNM)
 phenomenological model \cite{tool,narayanGardon,moynihan}.

We perform NVT molecular dynamics (MD) simulations for a system of
$N=216$ molecules, with periodic boundary conditions. Interactions are
cut off at a distance of $r=2.5 \sigma$, where $\sigma$ is the length
parameter defined in the SPC/E potential, and the reaction field method
is implemented to account for the long range interactions. We average
quantities over 32 independent trajectories at fixed density 
$\rho=1$~g/cm$^3$.  During cooling or heating, $T$ is continuously changed by
$\delta T=q \delta t$, where $q$ is the cooling/heating rate, and
$\delta t = 1$~fs is the elementary time step.  We perform: (i) cooling
scans at constant cooling rate down to $T=0$~K, starting from equilibrium
liquid configurations at $T=300$~K, (ii) heating scans at constant
heating rate, starting from $T=0$~K glass configurations, (iii) aging at
constant $T_{\rm age} =100$~K, where significant aging effects are
observed.  We study two cooling rates $q_c=-3 \times 10^{10}$~K/s and
$q_c=-10^{13}$~K/s, to mimic respectively the standard and hyperquenched
cooling rates, and one heating rate $q_h=+3 \times 10^{10}$~K/s.  Slow
experimental scan rates are typically $\approx 0.3$~K/s, while the
slowest simulation scan rate compatible with present computational
facilities is $10^7$ times faster ($\approx 3\times10^{10}$~K/s). 
Hence, the
temperature at which the system will lose equilibrium on cooling will be
significantly higher in simulations than in experiment.  Still, the key
fact that the structural relaxation time becomes longer than the
experimental (or simulation) time is the same for experiments and
simulations. Therefore, as we will show below, while
 the $T_g$ estimates differ, the $T$-dependence
and the phenomenology do not depend significantly on the scan rate.  In
hyperquench experiments, a cooling rate $10^4$ times faster than the
slow or `standard' rate is usually achieved, while in the present
simulations the faster quench rate is approximately 300 times faster than
 the slower quench rate.

\begin{figure}[htb]
\centerline {
\includegraphics[width=2.8in]{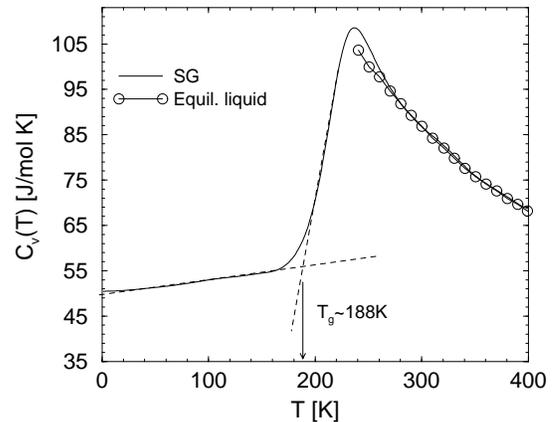}
}

\caption{Specific heat from MD simulations calculated by differentiating
 the total energy during heating of the standard glass (SG).  Circles
 denote equilibrium values of $C_V(T)$ in the liquid state.}
\label{e-Cv} 
\end{figure}

Figure \ref{e-Cv} shows the specific heat $C_V(T)$ calculated by
differentiating the temperature dependence of the total energy of the
system on heating at the rate $q_h=+3\times10^{10}$~K/s.  The glass
configurations are obtained by cooling equilibrium $T=300$~K liquid
configurations at the `standard' cooling rate $q_c=-3\times10^{10}$~K/s.
Following the usual experimental protocol, we estimate $T_g$ from the
intersection of the two dashed lines in Fig.\ref{e-Cv}.  The resulting
value, $T_g=188 K$, is slightly below the lowest $T$ at which
equilibrium simulations can be performed for
SPC/E\cite{francislong,galloprl}.  The $C_V(T)$ rise of
$\approx 55$~J/mol/K, is more than an order of magnitude 
larger than the experimentally
measured rise of $\approx 1.6-1.9$~J/mol/K\cite{johariWaterAB}.  For $T
\gtrsim 240$~K, $C_V(T)$ coincides with equilibrium data for the SPC/E
potential\cite{poole}. Indeed, the equilibrium relaxation time of the
system for $T \gtrsim 240$~K is $\lesssim 20$~ps, smaller than the
characteristic scan time $1$~K$/q_h \approx 30$~ps.

\begin{figure}[htb]
\centerline{
\includegraphics[width=2.8in]{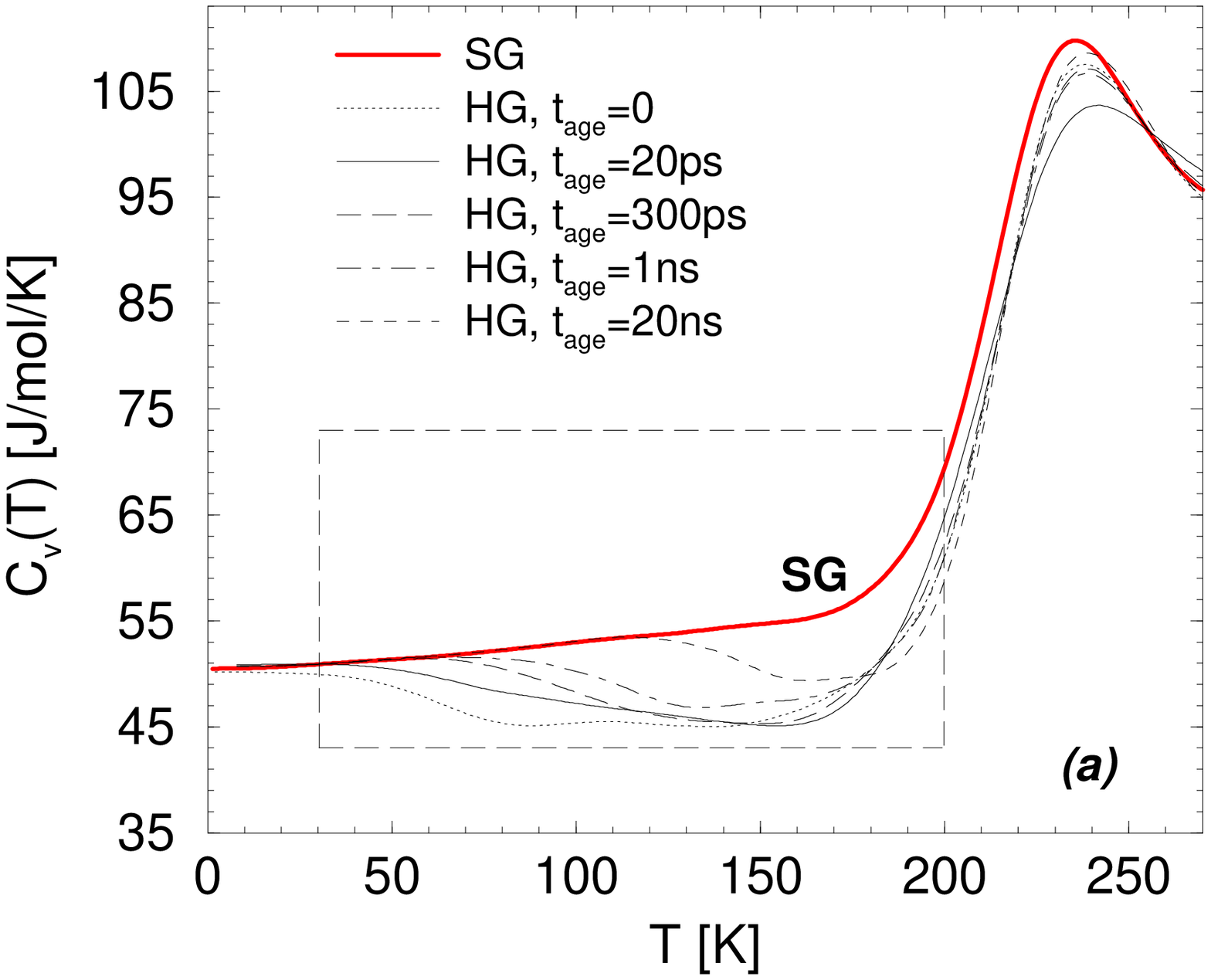}
}
\centerline {
\includegraphics[width=2.8in]{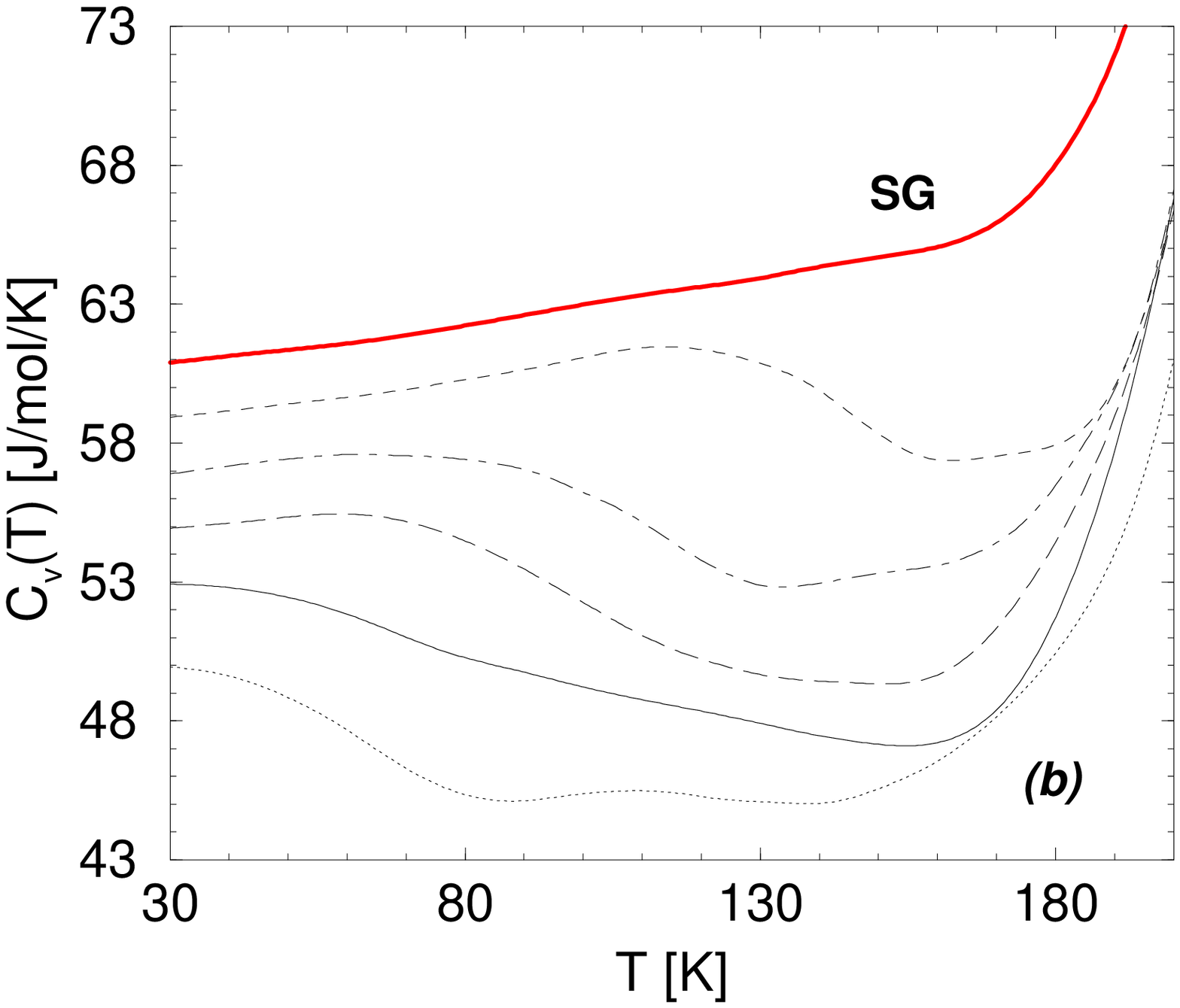}
}

\centerline {
\includegraphics[width=2.8in]{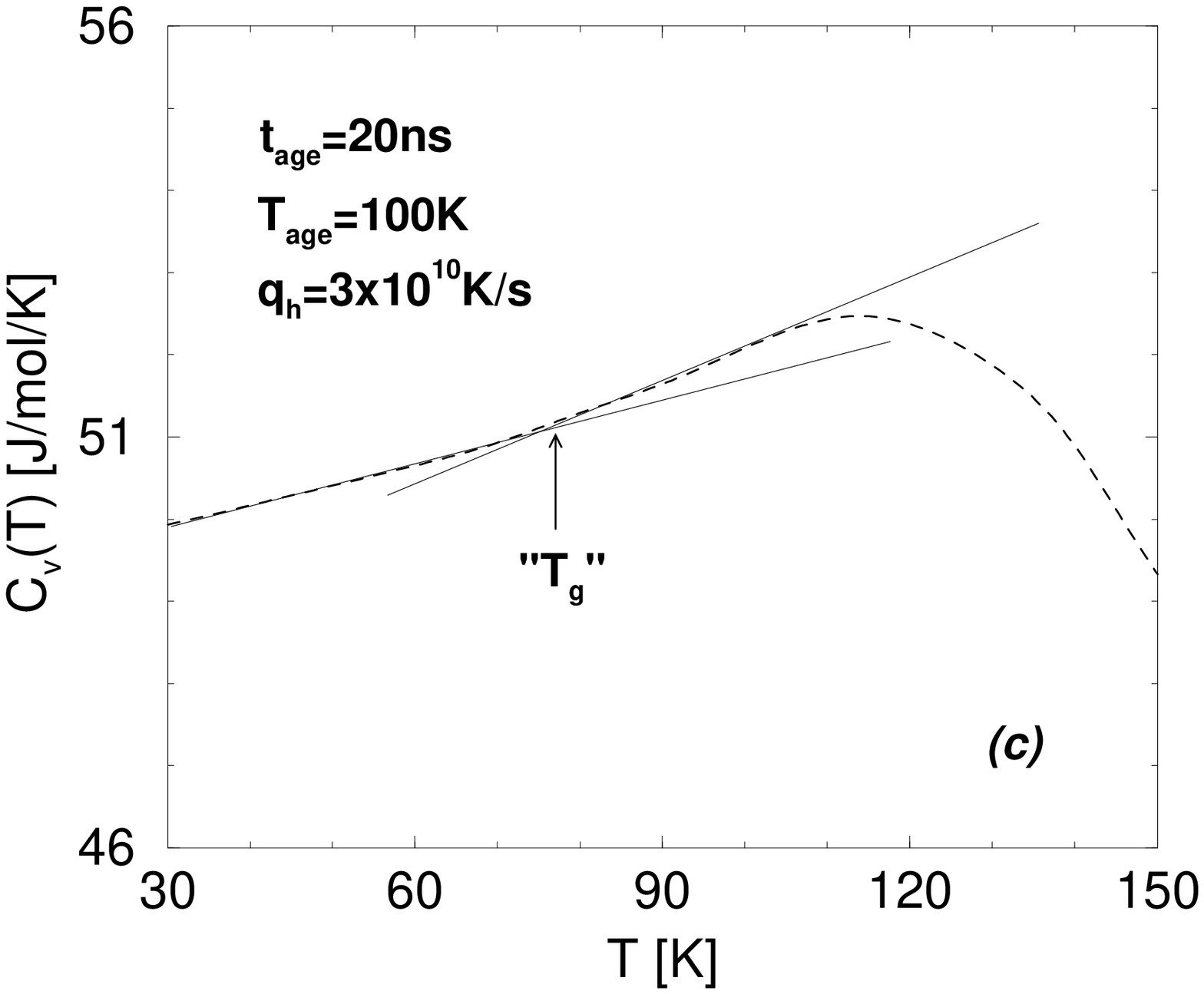}
}

\caption{(a) $T$-dependence of $C_V$ for heating scans of the standard
glass (SG) and the hyperquenched glass (HG).  Also shown are the heating
scans of the HG which has been annealed at $T_{age}=100$~K for four
different aging times $t_{age}$.  (b) Magnification of (a). Curves are
shifted for clarity by $3$~J/mol/K (for $t_{age}=20$~ps), $5$~J/mol/K
(for $t_{age}=300$~ps), $7$~J/mol/K (for $t_{age}=1$~ns), $9$~J/mol/K
(for $t_{age}=20$~ns), and $11$~J/mol/K (for SG). (c) Magnification of
$C_V(T)$ for the hyperquenched glass annealed at $T_{age}=100$~K for
$t_{age}=20$~ns, to highlight the weak pre-peak at $T \approx 113$
K. The straight lines show a possible construction that, in the absence
of the peak at $T \approx 220$~K, could be interpreted as the glass
transition temperature ``$T_g$''.}
\label{Cv-100K-zoom}
\end{figure}

\begin{figure}[htb]
\centerline{
\includegraphics[width=2.8in]{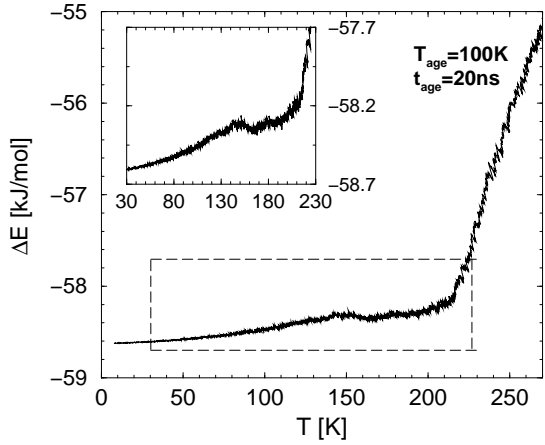}
}

\caption{$T$ dependence of $\Delta E \equiv E- 6RT$, where $E$ is the
total energy per molecule, $R$ is the gas constant, and $6RT$ is the
total energy of a glass of rigid molecules in the harmonic
approximation, contributing a constant $6R$ to $C_V$.  Using $\Delta E$
instead of $E$ amplifies the very weak signal, whose derivative is
responsible for the weak peak in $C_V$ shown in
Fig.\ref{Cv-100K-zoom}(c).}
\label{Cv-XXXK}
\end{figure}

We next compare in Figs.~\ref{Cv-100K-zoom}(a) and~\ref{Cv-100K-zoom}(b)
 the behavior of $C_V(T)$ on heating two different glasses, the
 `standard glass' obtained with the cooling rate $q_c=-3\times
 10^{10}$~K/s, and the `hyperquenched glass' obtained with the faster
 rate $q_c=-10^{13}$~K/s.  For the hyperquenched glass, $C_V(T)$ develops a
 valley for $T<T_g$, in agreement with DSC heating scan
 experiments\cite{angell,hodge2,gupta,yue}
 (indeed, Fig.\ref{Cv-100K-zoom}(a) is remarkably similar to Fig.~1 of
 Ref.\cite{gupta}).  The presence of a valley can be related to the
 descent of the system on the potential energy landscape upon heating
 with a rate slower than the cooling rate\cite{ourBriefComm}.
  Figs.~\ref{Cv-100K-zoom}(a) and~\ref{Cv-100K-zoom}(b) show $C_V(T)$ for
 the heating scan of the hyperquenched glass which has been annealed at
 $T_{\rm age} =100 K$, for different aging times $0 < t_{age} \leq 20$~ns.
 This annealing procedure is intended to mimic the experimental
 annealing procedure\cite{hodgeSumm}. Note that aging reduces the valley
 in $C_V(T)$, and that as $t_{age}$ increases, $C_V(T)$ evolves towards
 the standard glass value (Fig.~\ref{e-Cv}). Inspection of the curves
 for large $t_{age}$ (Fig.\ref{Cv-100K-zoom}(c)) shows that a small
 pre-peak appears at $T \approx 113$~K. If the standard protocol
 (Fig. 1) for the identification of the glass transition in the specific
 heat is applied, a $T_g$ of $\approx 75$~K is derived
 (Fig.\ref{Cv-100K-zoom}(c)). The amplitude of the peak in $C_V(T)$ is
 of the order of $1-2.5$~J/mol/K, and is reminiscent of the
 experimental value obtained in DSC measurements of hyperquenched water
 after annealing.  In the present case, in which crystallization does
 not interfere with the heating scan, there is no ambiguity in
 associating this peak with a precursor of the true glass transition
 which takes place at a much higher $T$.  To prove that the weak $C_V$
 pre-peak is outside the noise level, we show in Fig.~\ref{Cv-XXXK} the
 $T$-dependence of the energy, where a clear maximum can be observed.

Results presented in Figs.~\ref{Cv-100K-zoom} and~\ref{Cv-XXXK} are
 consistent with recent anneal-and-scan experiments \cite{angell} on
 hyperquenched inorganic glass which does not crystallize on heating.  Our
 simulations thus suggest that the measured specific heat peak
 (Ref.~\cite{Johari2}) which has been used to identify $T_g$ is a
 pre-peak associated with the use of a hyperquenched sample combined
 with the annealing procedure.
  
 Next, we discuss the possibility of modeling the simulation results
using the TNM approach\cite{moynihan,hodge2,hodgeSumm}, which is able to
model the experimental heating scan of the specific heat for glasses
generated with standard cooling rates (although it fails when applied to
hyperquenched glasses\cite{gupta}). The TNM model assumes the response
function of the system can be represented by a stretched exponential
function with stretching parameter $\beta$. It also assumes that the
relaxation time $\tau$ depends not only on the bath temperature $T$ but
also on a fictive temperature $T_f$ which accounts for the
out-of-equilibrium condition.  Narayanaswamy proposed that $\tau$ is
related to $T_f$ by
\begin{eqnarray} \tau(T,T_f) = A \exp \left[ \frac{x \Delta h^*}{R T} + \frac{ (1-x) \Delta h^*}{R T_f} \right] 
\label{Naray} 
\end{eqnarray} where $0 \leq x \leq 1$, $\Delta h^*$ and $A$ are constants
and $R$ is the ideal gas constant \cite{footnote}. In fact, Eq.~(\ref{Naray})
corresponds to Narayanaswamy's original equation rewritten by Moynihan
\cite{moynihan} who introduced the parameter $x$.
An alternative relation\cite{scherer} is offered by the
 generalization of the Adam-Gibbs expression, which connects, in
 equilibrium, $\tau$ to the configurational entropy $S_c$.  The
 resulting Adam-Gibbs-Scherer (AGS) expression is
\begin{eqnarray}
\tau(T,T_f) = A' \exp \left[ \frac{E_A}{S_c(T_f) T} \right]
\label{scherer}
\end{eqnarray}
where
\begin{eqnarray}
S_c(T)= \int_{T_K}^T \frac{\Delta C}{T} dT,
\end{eqnarray}
 $T_K$ is the Kauzmann temperature, $E_A$ is a constant, and $\Delta C$
 is the difference between the specific heats of the liquid and the
 glass.  The TNM model requires $\beta$ as a fitting parameter;
 additionally it requires the parameters ($A$, $x$, $\Delta h^*$) for the
 Narayanaswamy-expression, or ($A'$, $E_A$, $T_K$) for the AGS
 expression.
 
\begin{figure}[htb]
\centerline {
\includegraphics[width=2.8in]{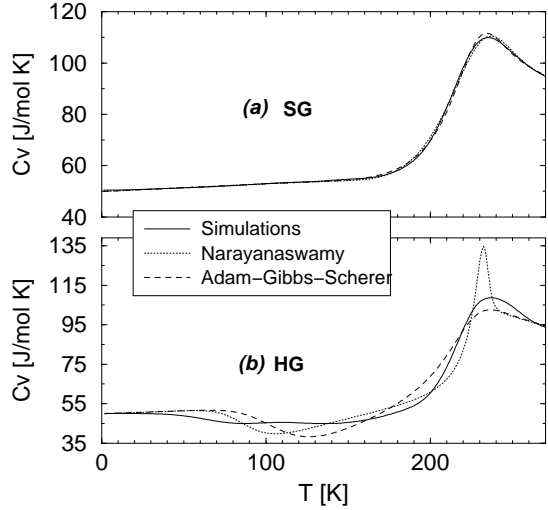}
}

\caption{Heating scans from the (a) standard glass (SG) and (b)
hyperquenched glass (HG). Our simulations are compared with the
predictions of the TNM model using both the Narayanaswamy model (fitting
parameters for the SG are $\ln(A/ns)=-22.36$, $\beta=0.525$, $x=0.635$,
and $\Delta h^*/R=4632$~K) and the Adam-Gibbs-Scherer expressions
(fitting parameters for the SG are $\ln(A/ns)=-9.86$, $\beta=0.519$, and
$E_A=27626$~kJ/mol; we use $S_c$ from Ref.~[32] and hence we
do not require $T_K$). We see that the TNM model describes the
behavior of $C_V(T)$ for the SG but fails for the HG.}
\label{Cv-30+30}   
\end{figure}

Figure~\ref{Cv-30+30} compares our MD results during the heating scan of
the standard and hyperquenched glass with the predictions of the TNM
model using both the Narayanaswamy and AGS expressions for $\tau$.  A
detailed analysis will be presented elsewhere\cite{futureTNGwork}; here
we show that both the Narayanaswamy and AGS
expressions give satisfactory results only for the standard glass
(Fig.~\ref{Cv-30+30}(a)).  The quality of the fit for the
hyperquenched glass is unsatisfactory, as observed in the analysis of
experimental data for hyperquenched samples\cite{gupta}, suggesting that
in the hyperquenched experiments, the aging sample cannot be
connected to a liquid at a fictive temperature $T_f$\cite{ourBriefComm}.
It also suggests that the application of the TNM approach for testing
the shadow glass transition must be taken with caution\cite{Johari2}.

In summary, we show by a proper numerical protocol that the complex
 phenomenology of the glass transition can be reproduced in simulations,
 notwithstanding the large differences in cooling rates.  The TNM model
 which is able to describe the experimental specific heat for the
 standard cooling rate also describes the corresponding simulation
 results.  One advantage of simulations is to shed light on phenomena
 that occur outside the experimentally-accessible
 region\cite{NatureGene}.  Our simulations show that the glass
 transition is characterized by a large specific heat peak, and when
 hyperquenched samples are annealed, the glass peak is
 anticipated at lower $T$ by a pre-peak, with a much lower amplitude.  This result
 supports the recent reinterpretation of the existing experimental data
 \cite{angell}, which identifies $136$~K as the temperature of the
 pre-peak, and suggests $\approx 165$~K as the `true' glass transition
 temperature.

We thank the BU Computation Center
for CPU time, and NSF Grants CHE0096892 and DMR0082535, MIUR Cofin 2001, Firb
2002 for support.

\end{document}